\documentclass{ieeeaccess}
\usepackage{cite}
\usepackage{amsmath,amssymb,amsfonts}
\usepackage{algorithm}
\usepackage{algpseudocode}

\usepackage{graphicx}
\usepackage{textcomp}
\usepackage{multirow}
\usepackage{booktabs}
\graphicspath{{figures/}}

\def\BibTeX{{\rm B\kern-.05em{\sc i\kern-.025em b}\kern-.08em
    T\kern-.1667em\lower.7ex\hbox{E}\kern-.125emX}}
\begin{document}
\history{Date of publication xxxx 00, 0000, date of current version xxxx 00, 0000.}
\doi{10.1109/ACCESS.2017.DOI}

\title{One-step Iterative Estimation of Effective Atomic Number and Electron Density for Dual Energy CT}
\author{\uppercase{Qian Wang}\authorrefmark{1},
\uppercase{Huiqiao Xie}\authorrefmark{2},
\uppercase{Tonghe Wang}\authorrefmark{2},
\uppercase{Justin Roper}\authorrefmark{1},
\uppercase{Hao Gao}\authorrefmark{3},
\uppercase{Zhen Tian}\authorrefmark{4},
\uppercase{Xiangyang Tang}\authorrefmark{5},
\uppercase{Jeffrey D. Bradley}\authorrefmark{1},
\uppercase{Tian Liu}\authorrefmark{6}, and
\uppercase{Xiaofeng Yang}\authorrefmark{1}
}
\address[1]{Department of Radiation Oncology and Winship Cancer Institute, Emory University, Atlanta, GA, 30322, USA}
\address[2]{Department of Medical Physics, Memorial Sloan Kettering Cancer Center, New York, NY, USA}
\address[3]{Department of Radiation Oncology, University of Kansas Medical Center, Kansas City, MO, USA}
\address[4]{Department of Radiation \& Cellular Oncology, University of Chicago, Chicago, IL, 60637, USA}
\address[5]{Department of Radiology and Imaging Sciences and Winship Cancer Institute, Emory University, Atlanta, GA, 30322, USA}
\address[6]{Department of Radiation Oncology, Icahn School of Medicine at Mount Sinai, New York, NY, 10029 USA}

\tfootnote{This research was supported in part by the National Institutes of Health under Award Number R01CA272991.}

\markboth
{Qian Wang \headeretal: One-step Iterative Estimation for Dual Energy CT}
{Qian Wang \headeretal: One-step Iterative Estimation for Dual Energy CT}

\corresp{Corresponding author: Xiaofeng Yang (e-mail: xiaofeng.yang@emory.edu).}

\begin{abstract}
Dual-energy computed tomography (DECT) is a promising technology that has shown a number of clinical advantages over conventional X-ray CT, such as improved material identification, artifact suppression, etc. For proton therapy treatment planning, besides material-selective images, maps of effective atomic number (Z) and relative electron density to that of water ($\rho_e$) can also be achieved and further employed to improve stopping power ratio accuracy and reduce range uncertainty. In this work, we propose a one-step iterative estimation method, which employs multi-domain gradient $L_0$-norm minimization, for Z and $\rho_e$ maps reconstruction. The algorithm was implemented on GPU to accelerate the predictive procedure and to support potential real-time adaptive treatment planning. The performance of the proposed method is demonstrated via both phantom and patient studies.
\end{abstract}


\titlepgskip=-15pt

\maketitle

\section{Introduction}
\label{sec:introduction}
\PARstart{C}{ompared} with traditional radiation therapy, proton therapy is superior in treatment of complex and recurrent tumors with maximized normal tissue and organ at risk (OAR) sparing beyond the target. This can effectively reduce the radiation toxicity and side effects \cite{zhang2010intensity}. However, proton therapy suffers from the range uncertainty issue and requires accurate dose calculation for treatment planning \cite{paganetti2012range}. Conventionally, either analytical or Monte Carlo dose calculation algorithms employs the stopping power ratio (SPR) or mass density that is obtained with the stoichiometric relationships from the Hounsfield units (HU) of the single energy simulation CT \cite{schneider1996calibration,schneider2000correlation,chang2020standardized,harms2020cone}. This category of methods relies on linear fitting of the SPR and mass density as a function of the planning CT HU numbers. It is susceptible to noise and fitting errors. In addition, HU ambiguity, i.e., in fact that different materials can have the same HU value is another issue in determination of the SPR and mass density. Therefore, a 3\% margin is typically reserved for the uncertainty of calculated proton range \cite{paganetti2012range}.

Dual-energy computed tomography (DECT) has been employed to avoid HU ambiguity and improve SPR accuracy \cite{bar2017potential,zhu2016dosimetric,yang2010theoretical}. In general, a two-step workflow is followed: material decomposition and generating maps of effective atomic number (Z) and relative electron density to that of water ($\rho_e$). With Z and $\rho_e$, SPR can be further calculated \cite{bourque2014stoichiometric,bar2018experimental,xie2018ex}. Material decomposition of DECT is typically an ill-posed inverse problem. During the material decomposition calculation process, the noise level can be substantially magnified, which further affects both spatial resolution and contrast. As a consequence, fine structures and low-contrast details may be submerged by noise, on top of the decreased accuracy of material decomposition. To solve this problem, image based, projection based, and iterative methods have been proposed. For image-based methods, decomposition is performed in reconstructed image domain by following a linear pattern \cite{Maas2009}. Projection based methods requires geometrical consistency and performs decomposition before reconstruction, i.e., in projection domain \cite{flohr2006first,stenner2007}. Iterative methods are proposed based on statistical models and nonlinear optimizations \cite{elbakri2002statistical,xu2009implementation,maass2009dual,niu2014iterative}. By introducing nonlinear relationship and smoothness mechanism, both beam-hardening artifacts and noise can be effectively suppressed. For the second step,  the calculation of Z and $\rho_e$ is an independent close-form process, which is separated from material decomposition \cite{alvarez1976energy,lehmann1981generalized,mei2018dual}. Thus, the final results of Z and $\rho_e$ highly depend on the image quality of decomposed material images. Meanwhile, because of the ill-posedness and deficiency of smoothness mechanism, noise can be amplified again in the step of Z and $\rho_e$ generation.

One popular smoothness strategy belongs to compressed sensing, which searches for sparse description in some specific domain. The most typical representatives are Tikhonov regularization ( $L_2$-norm of image gradient) \cite{Tikhonov1977} and TV regularization ( $L_1$-norm of image gradient) \cite{Rudin1992,Rudin1994,osher2003image}. Different from Tikhonov and TV regularization penalizing gradient magnitude, gradient $L_0$-norm punishes the gradient existence, which has the strongest capability of sparsity representation and is superior in edge preservation and noise suppression. However, the counting function of non-zero gradient leads to an NP-hard problem, which dramatically increases the solution difficulty \cite{blumensath2007iterative,peharz2012sparse}. To improve the feasibility, an approximation method was proposed by Xu et al \cite{Xu2011,xu2013unnatural}.

In this work, we develop an iterative one-step decomposition method with gradient $L_0$-norm minimization to estimate the effective atomic number and electron density from dual energy CT. Different from the aforementioned two-step approach, the established optimization model can obtain material composition, Z and $\rho_e$ maps simultaneously. Moreover, we introduce multi-domain regularizers for each searched-for variables. Thus, during the alternative solution process, the noise magnification problem can be consistently suppressed. Especially, gradient $L_0$-norm is superior to traditional Tikhonov and TV regularizations in sparse representation and edge preservation, which further benefits noise reduction and fidelity maintenance. The effectiveness and superior performance of the proposed method is demonstrated with both phantom and patient studies.

The remainder of this paper is organized as follows. In section \ref{sec:Method}, we review monoenergetic CT synthesization, Z and $\rho_e$ estimation, and gradient $L_0$-norm minimization. Then, we propose a multi-domain regularization based optimization model, and provide the algorithm for one-step estimation of material composition, Z and $\rho_e$ maps. In section \ref{sec:Results}, both phantom and patient studies are performed to verify the effectiveness of the proposed methods. In last section, we discuss some related issues and conclude this paper.

\section{Methods}\label{sec:Method}
In this section, the mathematical models of DECT material decomposition and monoemergetic CT synthesization are presented, and the estimation of Z and $\rho_e$ maps and solution process of gradient $L_0$ norm minimization are briefly reviewed. Then, by incorporating multi-domain gradient $L_0$ norm minimization, we establish an optimization model with multi-objectives. We also derive the iterative solution method and summarize the algorithm.

\subsection{Synthesis of monoenergetic CT image}
The basic assumption of DECT material decomposition is the linear attenuation coefficient ($\mu$) dependence on energy and material can be separated. Typically, there are two explanations, i.e., basis material decomposition and photoelectric/Compton effect based decomposition. For the first scenario, let $f_1$ and $f_2$ represent the composition images of two basis materials, $E_L$ and $E_H$ low and high energies employed for clinic DECT scan, and $\mu_1(\cdot)$ and $\mu_2(\cdot)$ the corresponding linear attenuation coefficients of the basis materials $f_1$ and $f_2$ under some specific energy setting. Thus, the achieved linear attenuation coefficient images of DECT can be described as,
\begin{subequations}
	\begin{align}
	\label{eq:lac_EL}
	\mu(E_L)=f_1\mu_1(E_L)+f_2\mu_2(E_L),\\
	\label{eq:lac_EH}
	\mu(E_H)=f_1\mu_1(E_H)+f_2\mu_2(E_H).
	\end{align}
\end{subequations}

After solving $f_1$ and $f_2$ from \eqref{eq:lac_EL} and \eqref{eq:lac_EH}, we can synthesize mass attenuation coefficient ($\mu/\rho$) image under some preferred monochromatic energy, such as $E_1$ and $E_2$ as follows,
\begin{subequations}
	\begin{align}
	\label{eq:mac_E1}
	\frac{\mu(E_1)}{\rho}=f_1\frac{\mu_1(E_1)}{\rho_1}+f_2\frac{\mu_2(E_1)}{\rho_2},\\
	\label{eq:mac_E2}
	\frac{\mu(E_2)}{\rho}=f_1\frac{\mu_1(E_2)}{\rho_1}+f_2\frac{\mu_2(E_2)}{\rho_2},
	\end{align}
\end{subequations}
where $\rho_1$ and $\rho_2$ are the mass density of the two basis materials. In this study, we use energy levels at 50 keV $(E_1)$ and 200 keV $(E_2)$, at which the photoelectric effect and the Compton effect are respectively the dominant X-ray interactions with matter.

\subsection{Estimation of electron density and effective atomic number}
For the energy range used in clinical CT, coherent (Rayleigh) scattering can often be neglected for standard body tissues. Thus, the mass-attenuation coefficient can be attributed to photoelectric absorption and incoherent (Compton) scattering, i.e.,
\begin{equation*}
	\frac{\mu(E)}{\rho}=a_p\psi_p(E)+a_c\psi_c(E),
\end{equation*}
where $\psi_p(E)$ and $\psi_c(E)$ are the energy dependencies of photoelectric absorption and Compton scattering, and $a_p$ and $a_c$ the characteristic constants of the material. The photoelectric term can be approximated by
\begin{equation*}
	a_p\psi_p(E)\simeq\rho_eC_p\frac{Z^m}{E^n},
\end{equation*}
where $\rho_e$ is the electron density, $Z$ the effective atomic number, and constants $C_p=9.8\times 10^{-24}$, $m=3.8$ and $n=3.2$ \cite{heismann2003density,mei2018dual}. The Compton scattering part can be represented by electron
density and the total Klein-Nishina cross-section, i.e., $a_c=\rho_e$ and
\begin{align}
	\psi_c(E)=KN(\gamma)=C_0\Big\{\frac{1+\gamma}{\gamma^2}\big[\frac{2(1+\gamma)}{1+2\gamma}-\frac{1}{\gamma}\ln(1+2\gamma)\big]\nonumber\\
	+\frac{1}{2\gamma}\ln(1+2\gamma)-\frac{1+3\gamma}{(1+2\gamma)^2}\Big\},\nonumber
\end{align} 
where $\gamma=\frac{E}{510.975 keV}$, $C_0=2\pi r_0^2$ and $r_0=2.818\times 10^{-13} cm$ is the classical electron radius.
Thus, \eqref{eq:mac_E1} and \eqref{eq:mac_E2} can also be described as,
\begin{subequations}
	\begin{align}
	\label{eq:mac_E1_eff}
	\frac{\mu(E_1)}{\rho}=\rho_e\big(C_p\frac{Z^m}{E_1^n}+\psi_c(E_1)\big),\\
	\label{eq:mac_E2_eff}
	\frac{\mu(E_2)}{\rho}=\rho_e\big(C_p\frac{Z^m}{E_2^n}+\psi_c(E_2)\big).
	\end{align}
\end{subequations}
By rewriting \eqref{eq:mac_E1_eff} and \eqref{eq:mac_E2_eff} as follows,
\small
\begin{subequations}
	\begin{align}
	\label{eq:rho_e}
	\big(\psi_c(E_1)E_1^n-\psi_c(E_2)E_2^n\big)\rho_e=\frac{\mu(E_1)}{\rho}E_1^n-\frac{\mu(E_2)}{\rho}E_2^n,\\
	\label{eq:Z_m}
	C_p\big(\frac{\mu(E_1)}{\rho}\frac{1}{E_2^n}-\frac{\mu(E_2)}{\rho}\frac{1}{E_1^n}\big)Z^m=\frac{\mu(E_2)}{\rho}\psi_c(E_1)\nonumber\\
	-\frac{\mu(E_1)}{\rho}\psi_c(E_2),
	\end{align}
\end{subequations} 
\normalsize
we can easily obtain $Z^m$ and $\rho_e$.

\subsection{Gradient $L_0$ norm minimization}
Sparse optimization is widely used in image smoothness and noise reduction. One popular preserving regularization is total variation (TV), which penalizes large gradient magnitudes, possibly influencing contrast
during smoothing. Another powerful choice is $L_0$ gradient minimization. As a sparse gradient counting scheme, gradient $L_0$-norm minimization is superior to TV in sparse representation without sacrificing contrast magnitude. Assume $\tilde{g}$ is the obtained noisy image and $g$ is the searched-for noise free image. Thus, the optimization model with gradient $L_0$-norm minimization can be expressed as \cite{Xu2011,xu2013unnatural},
\begin{equation}
	\label{eq:L0}
	\min_g\Big\{\|g-\tilde{g}\|_{L_2}^2+\lambda C(g)\Big\},
\end{equation}
where $\mathcal{C}(g)=\#\{\textbf{r}\in\Omega_2\big||\partial_xg(\textbf{r})|+|\partial_yg(\textbf{r})|\neq0\}$ represents the gradient $L_0$-norm, $\Omega_2$ the 2D spatial image domain, $\lambda$ a weight directly controlling the smoothness. Noticing \eqref{eq:L0} is a non-convex problem, we introduce auxiliary variables $h_g, v_g$ and rewrite the optimization model as,
\begin{equation}
	\label{eq:L0_aux}
	\min_g\Big\{\|g-\tilde{g}\|_{L_2}^2+\lambda C(h_g,v_g)\Big\},
\end{equation}
s.t. $h_g=\partial_xg$, $v_g=\partial_yg$, where $\mathcal{C}(h_g,v_g)=\#\{\textbf{r}\in\Omega_2\big||h_g(\textbf{r})|+|v_g(\textbf{r})|\neq0\}$.
Then we relax the constraints and convert \eqref{eq:L0_aux} to a unconstrained version, i.e.,
\begin{align}
	\label{eq:L0_unconstraint}
	\min_{g,h_g,v_g}\Big\{\|g-\tilde{g}\|_{L_2}^2+\lambda C(h_g,v_g)\nonumber\\
	+\beta\big(\|\partial_xg-h_g\|_{L_2}^2+\|\partial_yg-v_g\|_{L_2}^2\big)\Big\},
\end{align}
where $\beta$ is a relaxation parameter controlling the similarity between variables $(h_g,v_g)$ and their corresponding gradients. Further, \eqref{eq:L0_unconstraint} is split to the following subproblems,
\begin{subequations}
\begin{align}
	\label{eq:L0_sub_g}
	&\min_{g}\Big\{\|g-\tilde{g}\|_{L_2}^2+\beta\big(\|\partial_xg-h_g\|_{L_2}^2+\|\partial_yg-v_g\|_{L_2}^2\big)\Big\},\\
	\label{eq:L0_sub_hv}
	&\min_{h_g,v_g}\Big\{\|\partial_xg-h_g\|_{L_2}^2+\|\partial_yg-v_g\|_{L_2}^2+\frac{\lambda}{\beta} \mathcal{C}(h_g,v_g)\Big\}.
\end{align} 
\end{subequations}
For subproblem \eqref{eq:L0_sub_g}, we can derive the closed-form solution as follow,
\begin{equation*}
	g=\mathcal{F}^{-1}\Big(\frac{\mathcal{F}(\tilde{g})+\beta\big(\mathcal{F}^*(\partial_x)\mathcal{F}(h_g)+\mathcal{F}^*(\partial_y)\mathcal{F}(v_g)\big)}{\mathcal{F}(1)+\beta\big(\mathcal{F}^*(\partial_x)\mathcal{F}(\partial_x)+\mathcal{F}^*(\partial_y)\mathcal{F}(\partial_y)\big)}\Big),
\end{equation*}
where $\mathcal{F}(\cdot)$ and $\mathcal{F}^{*}(\cdot)$ represent Fast Fourier Transform operator and its complex conjugate. Subproblem \eqref{eq:L0_sub_hv} can be spatially decomposed as $\sum_{\textbf{r}\in\Omega_2}\min_{h_g,v_g}\mathcal{E}(\textbf{r})$, where $\mathcal{E}(\textbf{r})=\Big\{\big(h_g(\textbf{r})-\partial_xg(\textbf{r})\big)^2+\big(v_g(\textbf{r})-\partial_yg(\textbf{r})\big)^2+\frac{\lambda}{\beta}H\big(|h_g(\textbf{r})|+|v_g(\textbf{r})|\big)\Big\}$ and $H\big(|h_g(\textbf{r})|+|v_g(\textbf{r})|\big)$ is a binary function returning 1 if $|h_g(\textbf{r})|+|v_g(\textbf{r})|\neq0$ and 0 otherwise. The cost function $\mathcal{E}(\textbf{r})$ reaches its minimum under the condition
\footnotesize
\begin{equation}
	\label{eq:sol_hv}
	\big(h_g(\textbf{r}),v_g(\textbf{r})\big)=\begin{cases}
		(0,0), & \big(\partial_xg(\textbf{r})\big)^2+\big(\partial_yg(\textbf{r})\big)^2\leq \frac{\lambda}{\beta},\\
		(\partial_xg(\textbf{r}),\partial_yg(\textbf{r})), & otherwise.\end{cases}
\end{equation}
\normalsize

\subsection{Optimization based one-step estimation of electron density and effective atomic number}
Inspired by the estimation method of electron density and effective atomic number from synthesized monoenergetic CT image, we propose an optimization model with gradient $L_0$-norm minimization as follow,
\small
\begin{eqnarray}
	\label{eq:obj}
	\min_{f_1,f_2,\atop \rho_e,Z^m}\Big\{\|f_1\mu_1(E_L)+f_2\mu_2(E_L)-\mu(E_L)\|_{L_2}^2\nonumber\\
	+\|f_1\mu_1(E_H)+f_2\mu_2(E_H)-\mu(E_H)\|_{L_2}^2\nonumber\\
	+\|\rho_e\big(C_p\frac{Z^m}{E_1^n}+\psi_c(E_1)\big)-
	(f_1\frac{\mu_1(E_1)}{\rho_1}+f_2\frac{\mu_2(E_1)}{\rho_2})\|_{L_2}^2\nonumber\\
	+\|\rho_e\big(C_p\frac{Z^m}{E_2^n}+\psi_c(E_2)\big)-
	(f_1\frac{\mu_1(E_2)}{\rho_1}+f_2\frac{\mu_2(E_2)}{\rho_2})\|_{L_2}^2\nonumber\\
	+\lambda\big(\sum_{i=1}^2\mathcal{C}(f_i)+\mathcal{C}(\rho_e)+\mathcal{C}(Z^m)\big)\Big\}.
\end{eqnarray}
\normalsize
By introducing auxiliary variables $h_g, v_g (g=f_1,f_2,\rho_e,Z^m)$ and rewrite the optimization model as,
\small
\begin{eqnarray}
	\label{eq:obj_aux}
	\min_{f_1,f_2,\atop \rho_e,Z^m}\Big\{\|f_1\mu_1(E_L)+f_2\mu_2(E_L)-\mu(E_L)\|_{L_2}^2\nonumber\\
	+\|f_1\mu_1(E_H)+f_2\mu_2(E_H)-\mu(E_H)\|_{L_2}^2\nonumber\\
	+\|\rho_e\big(C_p\frac{Z^m}{E_1^n}+\psi_c(E_1)\big)-
	(f_1\frac{\mu_1(E_1)}{\rho_1}+f_2\frac{\mu_2(E_1)}{\rho_2})\|_{L_2}^2\nonumber\\
	+\|\rho_e\big(C_p\frac{Z^m}{E_2^n}+\psi_c(E_2)\big)-
	(f_1\frac{\mu_1(E_2)}{\rho_1}+f_2\frac{\mu_2(E_2)}{\rho_2})\|_{L_2}^2\nonumber\\
	+\lambda\sum_{g=f_1,f_2,\atop\rho_e,Z^m}\mathcal{C}(h_g,v_g)\Big\},
\end{eqnarray}
\normalsize
s.t. $h_g=\partial_xg$, $v_g=\partial_yg (g=f_1,f_2,\rho_e,Z^m)$. By relaxing the constraints, \eqref{eq:obj_aux} is converted to a unconstrained version as follow,
\small
\begin{eqnarray}
	\label{eq:obj_unconstraint}
	\min_{g,h_g,v_g,\atop g=f_1,f_2,\rho_e,Z^m}\Big\{\|f_1\mu_1(E_L)+f_2\mu_2(E_L)-\mu(E_L)\|_{L_2}^2\nonumber\\
	+\|f_1\mu_1(E_H)+f_2\mu_2(E_H)-\mu(E_H)\|_{L_2}^2\nonumber\\
	+\|\rho_e\big(C_p\frac{Z^m}{E_1^n}+\psi_c(E_1)\big)-
	(f_1\frac{\mu_1(E_1)}{\rho_1}+f_2\frac{\mu_2(E_1)}{\rho_2})\|_{L_2}^2\nonumber\\
	+\|\rho_e\big(C_p\frac{Z^m}{E_2^n}+\psi_c(E_2)\big)-
	(f_1\frac{\mu_1(E_2)}{\rho_1}+f_2\frac{\mu_2(E_2)}{\rho_2})\|_{L_2}^2\nonumber\\
	+\beta\sum_{g=f_1,f_2,\atop\rho_e,Z^m}\big(\|\partial_xg-h_g\|_{L_2}^2+\|\partial_yg-v_g\|_{L_2}^2\big)\nonumber\\
	+\lambda\sum_{g=f_1,f_2,\atop\rho_e,Z^m}\mathcal{C}(h_g,v_g)\Big\}.
\end{eqnarray}
\normalsize
We split \eqref{eq:obj_unconstraint} to the following subproblems,
\begin{subequations}
\begin{align}
	\label{eq:obj_sub_f}
	&\min_{f_1,f_2}\Big\{\|f_1\mu_1(E_L)+f_2\mu_2(E_L)-\mu(E_L)\|_{L_2}^2\nonumber\\
	&+\|f_1\mu_1(E_H)+f_2\mu_2(E_H)-\mu(E_H)\|_{L_2}^2\nonumber\\
	&+\|
	(f_1\frac{\mu_1(E_1)}{\rho_1}+f_2\frac{\mu_2(E_1)}{\rho_2})-\rho_e\big(C_p\frac{Z^m}{E_1^n}+\psi_c(E_1)\big)\|_{L_2}^2\nonumber\\
	&+\|
	(f_1\frac{\mu_1(E_2)}{\rho_1}+f_2\frac{\mu_2(E_2)}{\rho_2})-\rho_e\big(C_p\frac{Z^m}{E_2^n}+\psi_c(E_2)\big)\|_{L_2}^2\nonumber\\
	&+\beta\sum_{g=f_1,f_2}\big(\|\partial_xg-h_g\|_{L_2}^2+\|\partial_yg-v_g\|_{L_2}^2\big)\Big\},\\
	\label{eq:obj_sub_rho}
	&\min_{\rho_e}\Big\{\|\big(\psi_c(E_1)E_1^n-\psi_c(E_2)E_2^n\big)\rho_e\nonumber\\
	&-\big(\frac{\mu(E_1)}{\rho}E_1^n-\frac{\mu(E_2)}{\rho}E_2^n\big)\|_{L_2}^2\nonumber\\
	&+\tilde{\beta}\big(\|\partial_x\rho_e-h_{\rho_e}\|_{L_2}^2+\|\partial_y\rho_e-v_{\rho_e}\|_{L_2}^2\big)\Big\},\\
	\label{eq:obj_sub_z}
	&\min_{Z^m}\Big\{\|C_p\big(\frac{\mu(E_1)}{\rho}\frac{1}{E_2^n}-\frac{\mu(E_2)}{\rho}\frac{1}{E_1^n}\big)Z^m\nonumber\\
	&-\big(\frac{\mu(E_2)}{\rho}\psi_c(E_1)-\frac{\mu(E_1)}{\rho}\psi_c(E_2)\big)\|_{L_2}^2\nonumber\\
	&+\tilde{\beta}\big(\|\partial_xZ^m-h_{Z^m}\|_{L_2}^2+\|\partial_yZ^m-v_{Z^m}\|_{L_2}^2\big)\Big\},\\
	\label{eq:obj_sub_hv}
	&\min_{h_g,v_g,\atop g=f_1,f_2,\rho_e,Z^m}\Big\{\frac{\lambda}{\beta}\mathcal{C}(h_g,v_g)\nonumber\\
	&+\big(\|\partial_xg-h_g\|_{L_2}^2+\|\partial_yg-v_g\|_{L_2}^2\big)\Big\}.
\end{align} 
\end{subequations}
From subproblem \eqref{eq:obj_sub_f}, we can derive the closed-form solutions as follows,
\footnotesize
\begin{eqnarray*}
	f_1=\mathcal{F}^{-1}\Bigg(\bigg\{\mathcal{F}\big\{\sum\limits_{i=L,H}\mu_1(E_i)\big(\mu(E_i)-f_2\mu_2(E_i)\big)\\
	+\sum\limits_{j=1}^{2}\frac{\mu_1(E_j)}{\rho_1}\big[\rho_e\big(C_p\frac{Z^m}{E_j^n}+\psi_c(E_j)\big)-f_2\frac{\mu_2(E_j)}{\rho_2}\big]\big\}
	\\+\beta\big(\mathcal{F}^{*}(\partial_x)\mathcal{F}(h_{f_1})+\mathcal{F}^{*}(\partial_y)\mathcal{F}(v_{f_1})\big)\bigg\}/\bigg\{\mathcal{F}\big[\sum\limits_{i=L,H}\mu_1^2(E_i)\\+\sum\limits_{j=1}^{2}\big(\frac{\mu_1(E_j)}{\rho_1}\big)^2\big]
	+\beta\big(\mathcal{F}^{*}(\partial_x)\mathcal{F}(\partial_x)+\mathcal{F}^{*}(\partial_y)\mathcal{F}(\partial_y)\big)\bigg\}\Bigg),
\end{eqnarray*}
\begin{eqnarray*}
	f_2=\mathcal{F}^{-1}\Bigg(\bigg\{\mathcal{F}\big\{\sum\limits_{i=L,H}\mu_2(E_i)\big(\mu(E_i)-f_1\mu_1(E_i)\big)\\
	+\sum\limits_{j=1}^{2}\frac{\mu_2(E_j)}{\rho_2}\big[	\rho_e\big(C_p\frac{Z^m}{E_j^n}+\psi_c(E_j)\big)-f_1\frac{\mu_1(E_j)}{\rho_1}\big]\big\}\\
	+\beta\big(\mathcal{F}^{*}(\partial_x)\mathcal{F}(h_{f_2})+\mathcal{F}^{*}(\partial_y)\mathcal{F}(v_{f_2})\big)\bigg\}/\bigg\{\mathcal{F}\big[\sum\limits_{i=L,H}\mu_2^2(E_i)\\+\sum\limits_{j=1}^{2}\big(\frac{\mu_2(E_j)}{\rho_2}\big)^2\big]
	+\beta\big(\mathcal{F}^{*}(\partial_x)\mathcal{F}(\partial_x)+\mathcal{F}^{*}(\partial_y)\mathcal{F}(\partial_y)\big)\bigg\}\Bigg),
\end{eqnarray*}
\normalsize
where $\mathcal{F}(\cdot)$ and $\mathcal{F}(\cdot)^{*}$ represent Fast Fourier Transform operator and its complex conjugate. It is noticeable that subproblems \eqref{eq:obj_sub_rho} and \eqref{eq:obj_sub_z} are based on the equivalent conversions of \eqref{eq:mac_E1_eff} and \eqref{eq:mac_E2_eff}, i.e., \eqref{eq:rho_e} and \eqref{eq:Z_m}. Thus, to eliminate the cost difference caused by the substitution, relaxation parameter $\beta$ is replaced by a variant $\tilde{\beta}$. Furthermore, electron density $\rho_e$ can be solved from subproblem \eqref{eq:obj_sub_rho} by
\footnotesize
\begin{eqnarray*}
	\rho_e=\mathcal{F}^{-1}\Bigg(\bigg\{\mathcal{F}\big[\big(\psi_c(E_1)E_1^n-\psi_c(E_2)E_2^n\big)\big(\frac{\mu(E_1)}{\rho}E_1^n-\frac{\mu(E_2)}{\rho}E_2^n\big)\big]\\
	+\tilde{\beta}\big(\mathcal{F}^{*}(\partial_x)\mathcal{F}(h_{\rho_e})+\mathcal{F}^{*}(\partial_y)\mathcal{F}(v_{\rho_e})\big)\bigg\}/\bigg\{\mathcal{F}\big(\psi_c(E_1)E_1^n\\-\psi_c(E_2)E_2^n\big)^2
	+\tilde{\beta}\big(\mathcal{F}^{*}(\partial_x)\mathcal{F}(\partial_x)+\mathcal{F}^{*}(\partial_y)\mathcal{F}(\partial_y)\big)\bigg\}\Bigg).
\end{eqnarray*}
\normalsize
And effective atomic number $Z$ is the m-th root of $Z^m$, which can be obtained from subproblem \eqref{eq:obj_sub_z} by solving the following quadratic model,
\footnotesize
\begin{eqnarray*}
	\min_{Z^m}\Big\|   \big\{C_p^2\big(\frac{\mu(E_1)}{\rho}\frac{1}{E_2^n}-\frac{\mu(E_2)}{\rho}\frac{1}{E_1^n}\big)^2+\tilde{\beta}\big(\partial_x^{*}\partial_x+\partial_y^{*}\partial_y\big)\big\}Z^m \\
	-\big\{\big[C_p\big(\frac{\mu(E_1)}{\rho}\frac{1}{E_2^n}-\frac{\mu(E_2)}{\rho}\frac{1}{E_1^n}\big)\big(\frac{\mu(E_2)}{\rho}\psi_c(E_1)-\frac{\mu(E_1)}{\rho}\psi_c(E_2)\big)\big]\\
	+\tilde{\beta}\big(\partial_x^{*}(h_{Z^m})+\partial_y^{*}\mathcal{F}(v_{Z^m})\big)\big\}\Big\|_{L_2}^2
\end{eqnarray*}
\normalsize
Subproblem \eqref{eq:obj_sub_hv} can be solved by the same strategy for subproblem \eqref{eq:L0_sub_hv}, i.e., by using \eqref{eq:sol_hv}. Combing all the derivations, the iterative algorithm is summarized in Alg. \ref{alg:decom}.
\begin{algorithm*}[ht]
	\caption{Iterative algorithm of optimization model \eqref{eq:obj}}
	\label{alg:decom}
	\begin{algorithmic}
		\Require images $\mu(E_i) (i=L,H)$, energy parameters $\psi_c(E_j)$ and $\mu_j(E_i) (i=L,H, j=1,2)$, mass attenuation coefficients $\frac{\mu_1(E_j)}{\rho_1}$ and $\frac{\mu_1(E_j)}{\rho_1} (j=1,2)$, smoothing weight $\lambda$, parameters $\beta^{(0)}, \beta^{(max)}$, and rates $\tau$ and $\kappa$.
		\Ensure $f_1$, $f_2$, $\rho_e$, and $Z^m$
		\State \hspace{-1.3em} \textbf{Initialization: } $f_1^{(0)}=\frac{\mu_2(E_H)\mu(E_L)-\mu_2(E_L)\mu(E_H)}{\mu_1(E_L)\mu_2(E_H)-\mu_1(E_H)\mu_2(E_L)}$, $f_2^{(0)}=\frac{\mu_1(E_L)\mu(E_H)-\mu_1(E_H)\mu(E_L)}{\mu_1(E_L)\mu_2(E_H)-\mu_1(E_H)\mu_2(E_L)}$,\\
		$\big(\frac{\mu(E_1)}{\rho}\big)^{(0)}=f_1^{(0)}\frac{\mu_1(E_1)}{\rho_1}+f_2^{(0)}\frac{\mu_2(E_1)}{\rho_2}$, $\big(\frac{\mu(E_2)}{\rho}\big)^{(0)}=f_1^{(0)}\frac{\mu_1(E_2)}{\rho_1}+f_2^{(0)}\frac{\mu_2(E_2)}{\rho_2}$,\\
		$\rho_e^{(0)}=\frac{\big(\frac{\mu(E_1)}{\rho}\big)^{(0)}E_1^n-\big(\frac{\mu(E_2)}{\rho}\big)^{(0)}E_2^n}{\big(\psi_c(E_1)E_1^n-\psi_c(E_2)E_2^n\big)}$,
		$(Z^m)^{(0)}=\frac{\big(\frac{\mu(E_2)}{\rho}\big)^{(0)}\psi_c(E_1)-\big(\frac{\mu(E_1)}{\rho}\big)^{(0)}\psi_c(E_2)}{C_p\big(\big(\frac{\mu(E_1)}{\rho}\big)^{(0)}\frac{1}{E_2^n}-\big(\frac{\mu(E_2)}{\rho}\big)^{(0)}\frac{1}{E_1^n}\big)}$, $\tilde{\beta}^{(0)}=\tau\beta^{(0)}$,
		$k=0$.
		\Repeat
		\State With $g^{(k)}$, solve for $h_g^{(k)}$ and $v_g^{(k)}$ $(g=f_1,f_2,\rho_e,Z^m)$ in \eqref{eq:sol_hv}.
		\State With $h_g^{(k)}$ and $v_g^{(k)}$, solve for $g^{(k+1)}$ $(g=f_1,f_2,\rho_e,Z^m)$ with:
		\scriptsize
		\State $f_1^{(k+1)}=\mathcal{F}^{-1}\Big(\big\{\mathcal{F}\big\{\sum\limits_{i=L,H}\mu_1(E_i)\big(\mu(E_i)-f_2^{(k)}\mu_2(E_i)\big)+\sum\limits_{j=1}^{2}\frac{\mu_1(E_j)}{\rho_1}\big[\rho_e^{(k)}\big(C_p\frac{(Z^m)^{(k)}}{E_j^n}+\psi_c(E_j)\big)-f_2^{(k)}\frac{\mu_2(E_j)}{\rho_2}\big]\big\}+\beta^{(k)}\big(\mathcal{F}(\partial_x)^{*}\mathcal{F}(h_{f_1}^{(k)})+\mathcal{F}(\partial_y)^{*}\mathcal{F}(v_{f_1}^{(k)})\big)\big\}/\big\{\mathcal{F}\big[\sum\limits_{i=L,H}\mu_1^2(E_i)+\sum\limits_{j=1}^{2}\big(\frac{\mu_1(E_j)}{\rho_1}\big)^2\big]+\beta^{(k)}\big(\mathcal{F}(\partial_x)^{*}\mathcal{F}(\partial_x)+\mathcal{F}(\partial_y)^{*}\mathcal{F}(\partial_y)\big)\big\}\Big)$
		\State $f_2^{(k+1)}=\mathcal{F}^{-1}\Big(\big\{\mathcal{F}\big\{\sum\limits_{i=L,H}\mu_2(E_i)\big(\mu(E_i)-f_1^{(k)}\mu_1(E_i)\big)+\sum\limits_{j=1}^{2}\frac{\mu_2(E_j)}{\rho_2}\big[	\rho_e^{(k)}\big(C_p\frac{(Z^m)^{(k)}}{E_j^n}+\psi_c(E_j)\big)-f_1^{(k)}\frac{\mu_1(E_j)}{\rho_1}\big]\big\}+\beta^{(k)}\big(\mathcal{F}(\partial_x)^{*}\mathcal{F}(h_{f_2}^{(k)})+\mathcal{F}(\partial_y)^{*}\mathcal{F}(v_{f_2}^{(k)})\big)\big\}/\big\{\mathcal{F}\big[\sum\limits_{i=L,H}\mu_2^2(E_i)+\sum\limits_{j=1}^{2}\big(\frac{\mu_2(E_j)}{\rho_2}\big)^2\big]+\beta^{(k)}\big(\mathcal{F}(\partial_x)^{*}\mathcal{F}(\partial_x)+\mathcal{F}(\partial_y)^{*}\mathcal{F}(\partial_y)\big)\big\}\Big)$
		\normalsize
		\State $\big(\frac{\mu(E_1)}{\rho}\big)^{(k+1)}=f_1^{(k)}\frac{\mu_1(E_1)}{\rho_1}+f_2^{(k)}\frac{\mu_2(E_1)}{\rho_2}$
		\State $\big(\frac{\mu(E_2)}{\rho}\big)^{(k+1)}=f_1^{(k)}\frac{\mu_1(E_2)}{\rho_1}+f_2^{(k)}\frac{\mu_2(E_2)}{\rho_2}$
		\scriptsize
		\State  $\rho_e^{(k+1)}=\mathcal{F}^{-1}\Big(\big\{\mathcal{F}\Big[\big(\psi_c(E_1)E_1^n-\psi_c(E_2)E_2^n\big)\Big(\big(\frac{\mu(E_1)}{\rho}\big)^{(k)}E_1^n-\big(\frac{\mu(E_2)}{\rho}\big)^{(k)}E_2^n\Big)\Big]+\tilde{\beta}^{(k)}\big(\mathcal{F}(\partial_x)^{*}\mathcal{F}(h_{\rho_e}^{(k)})+\mathcal{F}(\partial_y)^{*}\mathcal{F}(v_{\rho_e}^{(k)})\big)\big\}/\big\{\mathcal{F}\big(\psi_c(E_1)E_1^n-\psi_c(E_2)E_2^n\big)^2+\tilde{\beta}^{(k)}\big(\mathcal{F}(\partial_x)^{*}\mathcal{F}(\partial_x)+\mathcal{F}(\partial_y)^{*}\mathcal{F}(\partial_y)\big)\big\}\Big)$
		\State
		$\min\limits_{(Z^m)^{(k+1)}}\Big\|   \big\{C_p^2\Big(\big(\frac{\mu(E_1)}{\rho}\big)^{(k)}\frac{1}{E_2^n}-\big(\frac{\mu(E_2)}{\rho}\big)^{(k)}\frac{1}{E_1^n}\Big)^2+\tilde{\beta}^{(k)}\big(\partial_x^{*}\partial_x+\partial_y^{*}\partial_y\big)\big\}Z^m -\big\{\big[C_p\Big(\big(\frac{\mu(E_1)}{\rho}\big)^{(k)}\frac{1}{E_2^n}-\big(\frac{\mu(E_2)}{\rho}\big)^{(k)}\frac{1}{E_1^n}\Big)\Big(\big(\frac{\mu(E_2)}{\rho}\big)^{(k)}\psi_c(E_1)-\big(\frac{\mu(E_1)}{\rho}\big)^{(k)}\psi_c(E_2)\Big)\big]+\tilde{\beta}^{(k)}\big(\partial_x^{*}(h_{Z^m}^{(k)})+\partial_y^{*}\mathcal{F}(v_{Z^m}^{(k)})\big)\big\}\Big\|_{L_2}^2$
		\normalsize
		\State $\beta^{(k)}=\kappa\beta^{(k)}$
		\State $\tilde{\beta}^{(k)}=\tau\beta^{(k)}$
		\State  $k=k+1$
		\Until $\beta^{(k)}\geq\beta^{(max)}$
	\end{algorithmic}
\end{algorithm*}

\section{RESULTS}
\label{sec:Results}
\begin{figure}[ht]
	\centering
	\includegraphics[width=0.5\textwidth]{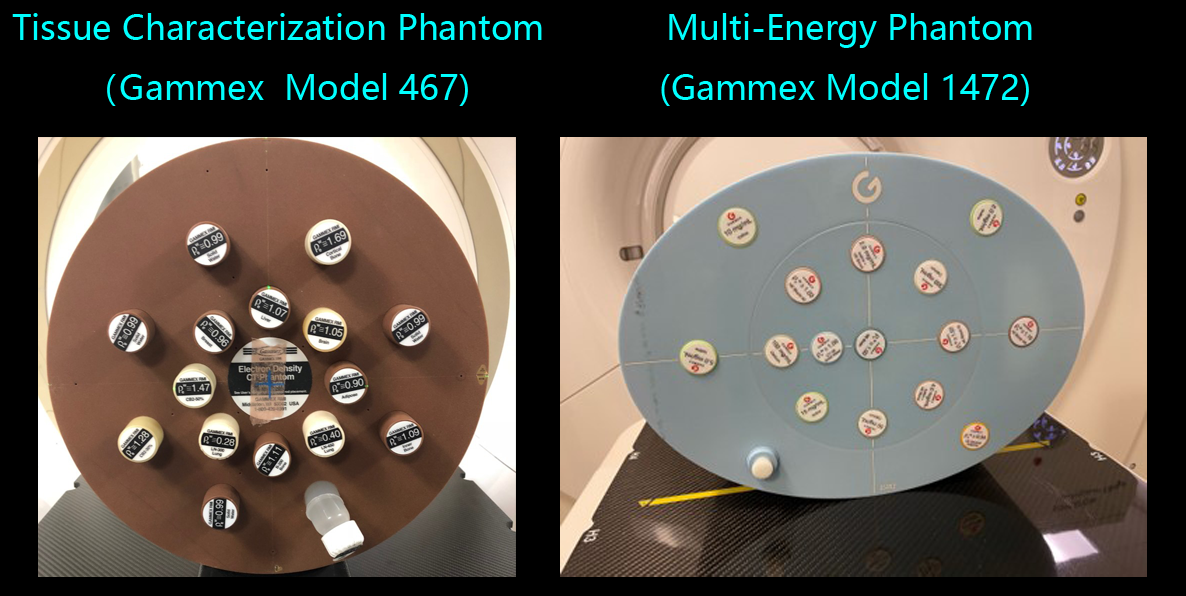}
	\caption{Phantom illustrations.}\label{fig:Phan}
\end{figure}

In this section, we applied the proposed method to two phantom studies and one patient study respectively. The employed phantoms are presented in Fig. \ref{fig:Phan}. All the data were collected by a Siemens SOMATOM Definition Edge scanner using TwinBeam protocols with 120 kV. The x-ray tube current is 210 mA for tissue characterization phantom, 655 mA for multi-energy phantom study and 261mA for patient study. Slice thickness is 0.5 mm with a reconstruction diameter of 500 mm. We compared the proposed method with direct two-step decomposition \cite{granton2008implementation} and estimation method \cite{mei2018dual}. For visual evaluations, we also magnified local patches and extracted line profiles for detail comparisons. Moreover, for phantom-based quantitative assessments., three numerical image quality measures are employed, i.e., peak signal-to-noise ratio (PSNR) \cite{huynh2008scope}, normal mean absolute deviation (NMAD) \cite{zhu2012noise} and structural similarity (SSIM) \cite{wang2004image}.

\subsection{Tissue characterization phantom study}
\begin{figure*}[ht]
	\centering
	\includegraphics[width=\textwidth]{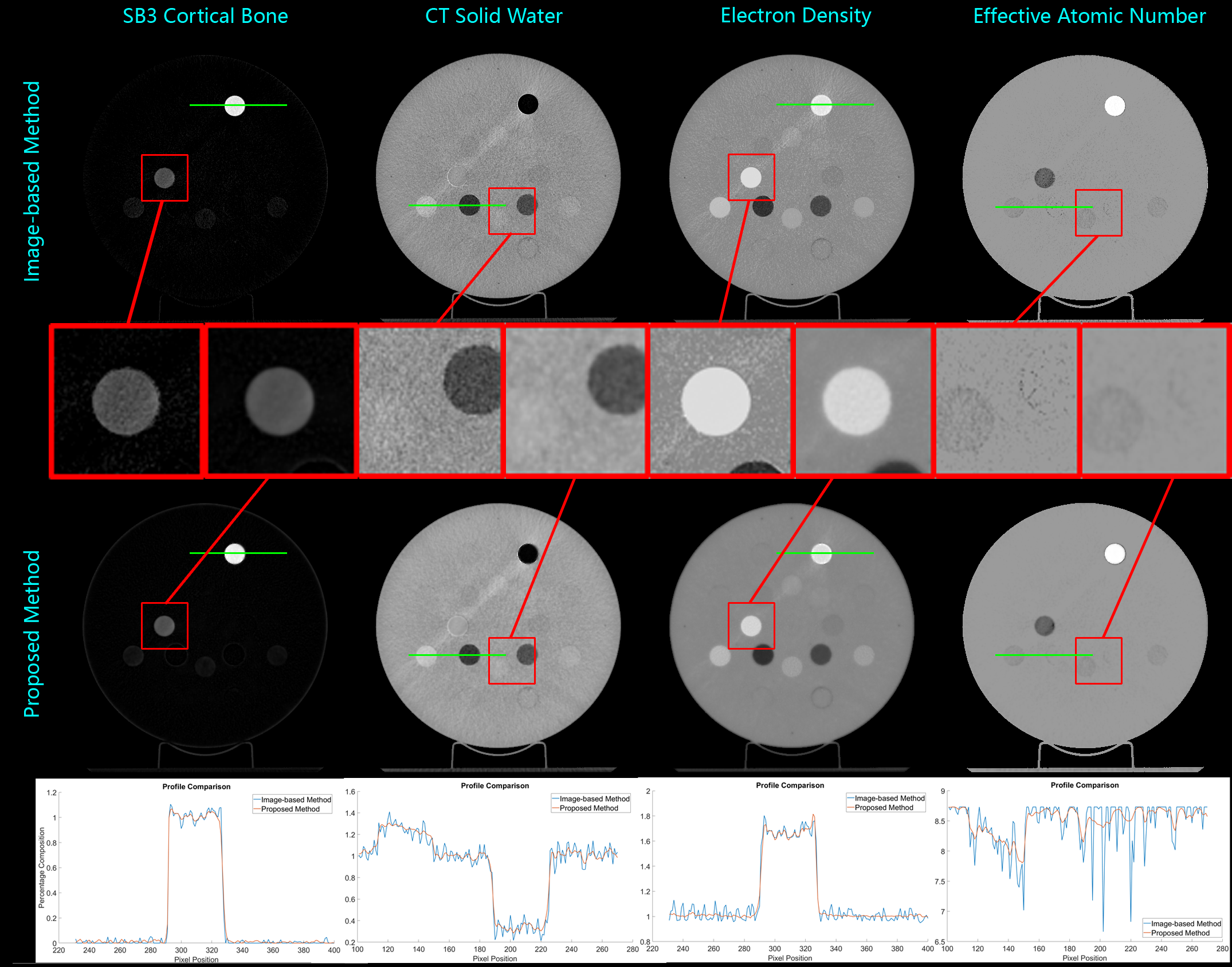}
	\caption{Material decomposition of tissue characterization phantom study by using classic image-based method and the proposed method.}\label{fig:Result_Phan}
\end{figure*}

First, the Gammex tissue characterization phantom model 467 (Gammex Inc., Middleton,WI) was scanned, which included 16 insert locations with human reference tissues. In this experiment, the 16 inserts consist of 13 materials. The material list and the reference valuse of $\rho_e$ and Z are shown in Table \ref{tbl:Phan}. The basis materials are chosen as SB3 cortical bone and CT solid water. The decomposed material images and maps of $\rho_e$ and Z are presented in Fig. \ref{fig:Result_Phan}. Results of estimation accuracy are provided in Table \ref{tbl:Phan}. Image quality assessments are shown in Table \ref{tbl:Phan_IQA}.

From Fig. \ref{fig:Result_Phan}, the zoomed-in local patches of the proposed method clearly demonstrated the superior performance in noise suppression and edge preservation, which also can be found from line profile comparisons. Moreover, the results of Table \ref{tbl:Phan_IQA} quantitatively verified the proposed method outperformed in image fidelity and smoothness. Noticeably, the proposed method accurately estimated $\rho_e$ and Z maps (see Table \ref{tbl:Phan}), of which the relative error is consistently smaller than 0.24 for $\rho_e$ map and 0.59 for Z map.

\subsection{Multi-energy phantom study}
\begin{figure*}[ht]
	\centering
	\includegraphics[width=\textwidth]{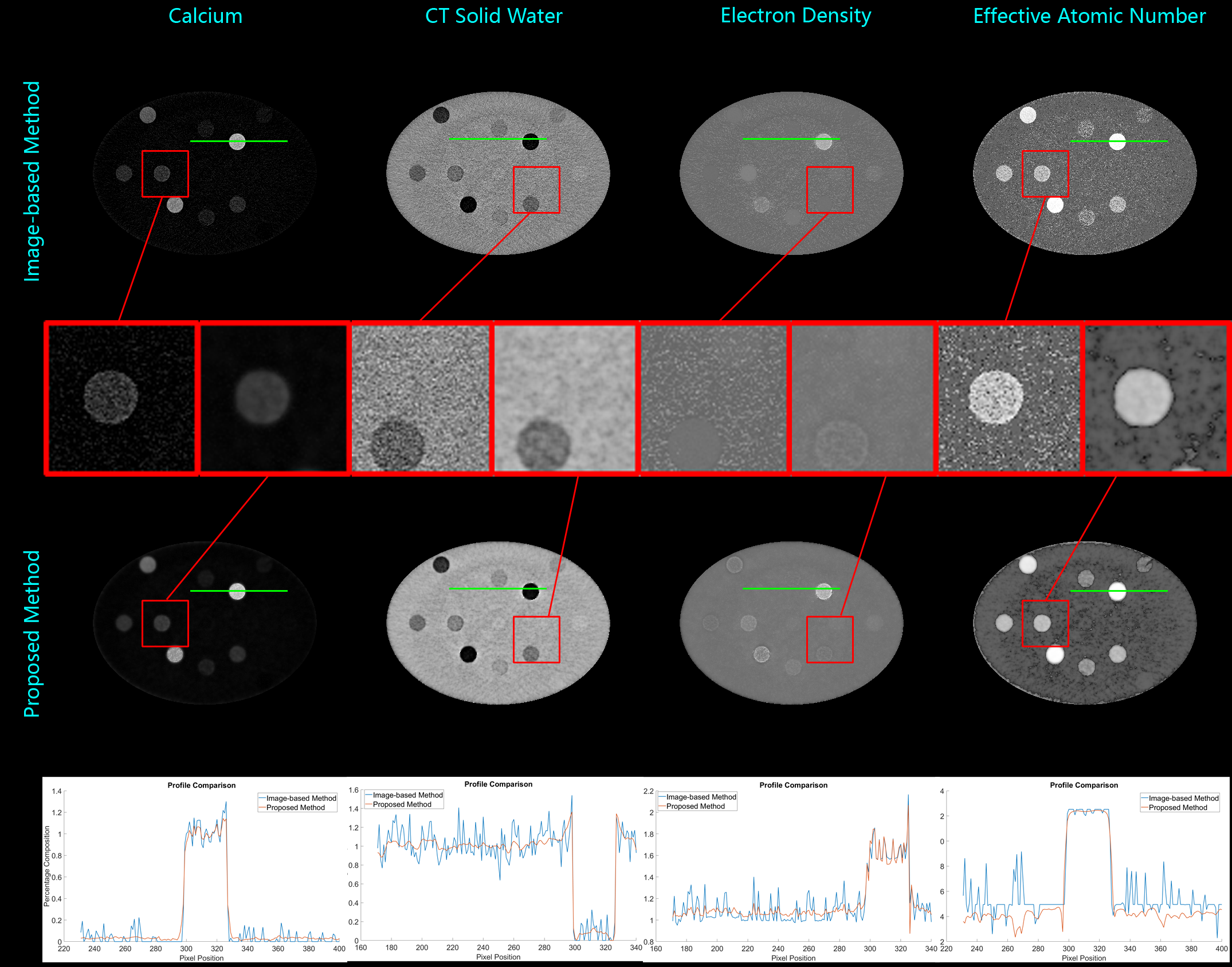}
	\caption{Material decomposition of multi-energy phantom study by using classic image-based method and the proposed method.}\label{fig:Result_Gammex}
\end{figure*}

We also scanned Gammex multi-energy phantom model 1472 (Gammex Inc., Middleton,WI). The phantom includes 15 inserts with various materials. The material list and the reference valuse of $\rho_e$ and Z are shown in Table \ref{tbl:Gammex}, where the reference Z valuse are calculated with Mayneord's power-law method ($Z=\sqrt[n]{\sum_i \alpha_iZ_i^n}$) with $n = 2.94$, where $Z_i$ is the atomic number of element $i$ and $\alpha_i$ is the fraction of the total number of electrons associated with element $i$ \cite{schaeffer2021accuracy}. Calcium and CT solid water were selected as basis materials. Fig. \ref{fig:Result_Gammex} illustrates the decomposed material images and maps of $\rho_e$ and Z. Estimation accuracy and image quality assessments are shown in Tables \ref{tbl:Gammex} and \ref{tbl:Gammex_IQA} respectively.

It is obvious, this study is more noisy than the previous one because of larger phantom size, but the proposed method stills works well for the severe scenario (see Fig. \ref{fig:Result_Gammex}). Moreover, the estimation accuracy of the proposed method is still reasonable. The relative error is consistently smaller than 0.27 for $\rho_e$ map and 0.48 for Z map. Both phantom studies consistently demonstrate the superiority of the proposed method in noise reduction, and accurate estimation.

\subsection{Patient study}
\begin{figure*}[ht]
	\centering
	\includegraphics[width=\textwidth]{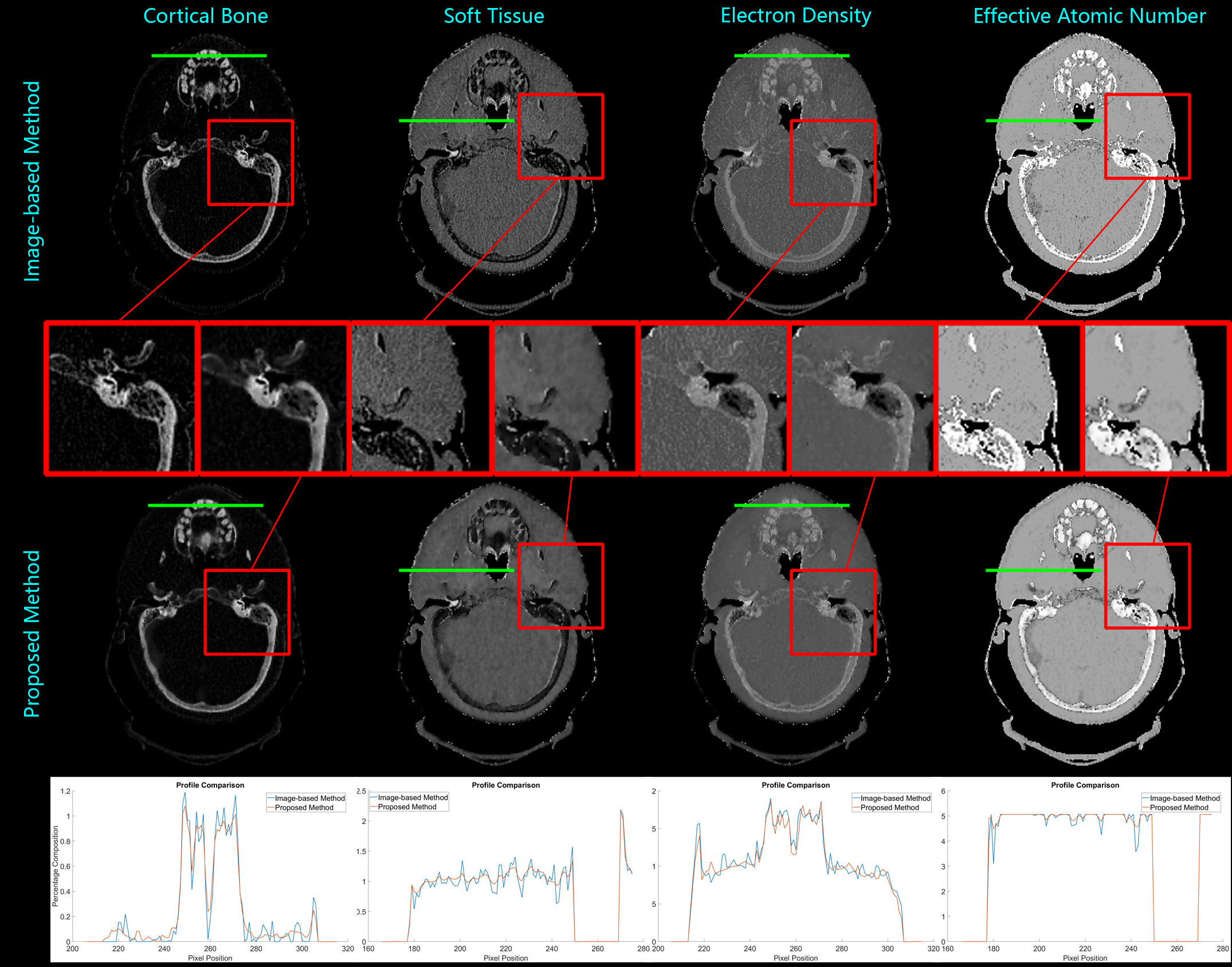}
	\caption{Material decomposition of patient study by using classic image-based method and the proposed method.}\label{fig:Result_Patient}
\end{figure*}

For the patient experiment shown in Fig. \ref{fig:Result_Patient},  the two basis materials are selected as cortical bone and soft tissue. From the bone comparisons, we can figure out the proposed method well preserves the edges and concentration, but the direct method performs inferior. From the soft tissue images, we can also find the direct decomposition image suffers from obvious noise, but the proposed method can effectively suppress it. The same conclusion can also be made from $\rho_e$ and Z comparisons.

\begin{table*}[htbp]
	\caption{Electron density and effective atomic number comparisons of tissue characterization phantom study.}\label{tbl:Phan}
	\begin{center}
	\footnotesize
	\begin{tabular}{rr|c|c|c|c||c|c|c|c}
		\toprule
		&&\multicolumn{4}{|c||}{Electron density}&\multicolumn{4}{c}{Effective atomic number}\\
		&&Reference&Mean&Standard&Relative&Reference&Mean&Standard&Relative\\
		&&Value&Value&Deviation&Error&Value&Value&Deviation&Error\\
		\midrule
		\multirow{2}{*}{SB3 Cortical Bone}&Image-based Method& \multirow{2}{*}{1.6900}&1.6686&\textbf{0.0454}&0.0127& \multirow{2}{*}{14.1400}&14.2982&\textbf{0.1256}&0.0112\\
		&Proposed Method& &\textbf{1.6754}&0.0489&\textbf{0.0086}& &\textbf{14.1683}&0.3024&\textbf{0.0020}\\
		\hline
		\multirow{2}{*}{CB2 - 50\% CaCO3}&Image-based Method& \multirow{2}{*}{1.4700}&1.5850&\textbf{0.0026}&0.0782& \multirow{2}{*}{12.9800}&5.2557&0.8418&0.5951\\
		&Proposed Method& &\textbf{1.5788}&0.0223&\textbf{0.0740}& &\textbf{5.4246}&\textbf{0.5667}&\textbf{0.5821}\\
		\hline
		\multirow{2}{*}{CB2 - 30\% CaCO3}&Image-based Method& \multirow{2}{*}{1.2800}&1.3637&\textbf{0.0150}&0.0654& \multirow{2}{*}{11.3900}&8.1450&0.3151&0.2849\\
		&Proposed Method& &\textbf{1.3524}&0.0242&\textbf{0.0566}& &\textbf{8.1684}&\textbf{0.1480}&\textbf{0.2828}\\
		\hline
		\multirow{2}{*}{B200 Bone Mineral}&Image-based Method& \multirow{2}{*}{1.1000}&1.1692&\textbf{0.0181}&0.0629& \multirow{2}{*}{10.9000}&8.1371&0.3812&0.2535\\
		&Proposed Method& &\textbf{1.1647}&\textbf{0.0181}&\textbf{0.0588}& &\textbf{8.1764}&\textbf{0.1573}&\textbf{0.2499}\\
		\hline
		\multirow{2}{*}{IB Inner Bone}&Image-based Method& \multirow{2}{*}{1.0900}&1.1599&0.0193&0.0641& \multirow{2}{*}{10.9000}&8.2477&0.3217&0.2433\\
		&Proposed Method& &\textbf{1.1550}&\textbf{0.0185}&\textbf{0.0596}& &\textbf{8.2766}&\textbf{0.1265}&\textbf{0.2407}\\
		\hline
		\multirow{2}{*}{LV1 Liver}&Image-based Method& \multirow{2}{*}{1.0600}&1.1224&0.0601&0.0588& \multirow{2}{*}{8.1100}&\textbf{8.6285}&0.1897&\textbf{0.0639}\\
		&Proposed Method& &\textbf{1.1080}&\textbf{0.0215}&\textbf{0.0453}& &8.6506&\textbf{0.0877}&0.0667\\
		\hline
		\multirow{2}{*}{BRN-SR2 Brain}&Image-based Method& \multirow{2}{*}{1.0400}&1.0633&0.0553&0.0224& \multirow{2}{*}{6.3100}&\textbf{8.6457}&0.1744&\textbf{0.3702}\\
		&Proposed Method& &\textbf{1.0520}&\textbf{0.0123}&\textbf{0.0116}& &8.6665&\textbf{0.0623}&0.3735\\
		\hline
		\multirow{2}{*}{True Water}&Image-based Method& \multirow{2}{*}{1.0000}&\textbf{0.9955}&0.0514&\textbf{0.0045}& \multirow{2}{*}{7.4200}&\textbf{8.6333}&0.2084&\textbf{0.1635}\\
		&Proposed Method& &0.9833&\textbf{0.0274}&0.0167& &8.6548&\textbf{0.0951}&0.1664\\
		\hline
		\multirow{2}{*}{CT Solid Water}&Image-based Method& \multirow{2}{*}{0.9900}&1.0125&0.0485&0.0228& \multirow{2}{*}{8.1100}&\textbf{8.6099}&0.2102&\textbf{0.0616}\\
		&Proposed Method& &\textbf{1.0053}&\textbf{0.0173}&\textbf{0.0155}& &8.6305&\textbf{0.0849}&0.0642\\
		\hline
		\multirow{2}{*}{BR-12 Breast}&Image-based Method& \multirow{2}{*}{0.9600}&0.9783&0.0481&0.0190& \multirow{2}{*}{7.2400}&\textbf{8.6044}&0.2092&\textbf{0.1885}\\
		&Proposed Method& &\textbf{0.9736}&\textbf{0.0109}&\textbf{0.0141}& &8.6268&\textbf{0.0721}&0.1915\\
		\hline
		\multirow{2}{*}{AP6 Adipose}&Image-based Method& \multirow{2}{*}{0.9300}&0.9554&0.0517&0.0273& \multirow{2}{*}{6.4000}&\textbf{8.6161}&0.2115&\textbf{0.3463}\\
		&Proposed Method& &\textbf{0.9478}&\textbf{0.0100}&\textbf{0.0192}& &8.6396&\textbf{0.0744}&0.3499\\
		\hline
		\multirow{2}{*}{LN-450 Lung}&Image-based Method& \multirow{2}{*}{0.4400}&\textbf{0.5191}&0.0495&\textbf{0.1798}& \multirow{2}{*}{7.8400}&\textbf{8.5520}&0.3933&\textbf{0.0908}\\
		&Proposed Method& &0.5278&\textbf{0.0312}&0.1995& &8.5893&\textbf{0.1117}&0.0956\\
		\hline
		\multirow{2}{*}{LN-300 Lung}&Image-based Method& \multirow{2}{*}{0.2900}&\textbf{0.3495}&0.0516&\textbf{0.2053}& \multirow{2}{*}{7.8600}&\textbf{8.5076}&0.5974&\textbf{0.0824}\\
		&Proposed Method& &0.3582&\textbf{0.0311}&0.2353& &8.5691&\textbf{0.1154}&0.0902\\
		\bottomrule
	\end{tabular}
	\normalsize
\end{center}
\end{table*}

\begin{table}[ht]
	\caption{Image quality assessments of tissue characterization phantom study.}\label{tbl:Phan_IQA}
	\begin{center}
		\begin{tabular}{r|c|c|c}
			\toprule\vspace{2pt}
			&\multicolumn{3}{c}{Electron Density}\vspace{2pt}\\
			&PSNR&NMAD&SSIM\\
			\midrule
			Image-based Method&39.6855&0.0514&0.9841\\
			Proposed Method&\textbf{42.0432}&\textbf{0.0383}&\textbf{0.9976}\\
			\bottomrule
			&\multicolumn{3}{c}{Effective Atomic Number}\vspace{2pt}\\
			&PSNR&NMAD&SSIM\\
			\midrule
			Image-based Method&27.5920&0.1660&0.9859\\
			Proposed Method&\textbf{27.7670}&\textbf{0.1645}&\textbf{0.9944}\\
			\bottomrule
		\end{tabular}
	\end{center}
\end{table}

\begin{table*}[ht]
	\caption{Electron density and effective atomic number comparisons of multi-energy phantom study.}\label{tbl:Gammex}
	\begin{center}
	\footnotesize
	\begin{tabular}{rr|c|c|c|c||c|c|c|c}
		\toprule
		&&\multicolumn{4}{|c||}{Electron density}&\multicolumn{4}{c}{Effective atomic number}\\
		&&Reference&Mean&Standard&Relative&Reference&Mean&Standard&Relative\\
		&&Value&Value&Deviation&Error&Value&Value&Deviation&Error\\
		\midrule
		\multirow{2}{*}{Calcium 300 mg/mL}&Image-based Method& \multirow{2}{*}{1.4500}&\textbf{1.6441}&\textbf{0.0991}&\textbf{0.1338}& \multirow{2}{*}{12.2100}&12.3824&0.2364&0.0141\\
		&Proposed Method& &1.6504&0.1117&0.1382& &\textbf{12.3724}&\textbf{0.0723}&\textbf{0.0133}\\
		\hline
		\multirow{2}{*}{Calcium 100mg/mL}&Image-based Method& \multirow{2}{*}{1.1900}&\textbf{1.2225}&\textbf{0.0144}&\textbf{0.0273}& \multirow{2}{*}{9.8700}&9.0083&1.3748&0.0873\\
		&Proposed Method& &1.2254&0.0461&0.0297& &\textbf{9.2112}&\textbf{0.3935}&\textbf{0.0667}\\
		\hline
		\multirow{2}{*}{Calcium 50mg/mL}&Image-based Method& \multirow{2}{*}{1.1300}&\textbf{1.1469}&\textbf{0.0268}&\textbf{0.0150}& \multirow{2}{*}{8.8000}&6.7705&2.1205&0.2306\\
		&Proposed Method& &1.1489&0.0390&0.0167& &\textbf{7.1021}&\textbf{0.8801}&\textbf{0.1929}\\
		\hline
		Iodine+HE Blood &Image-based Method& \multirow{2}{*}{1.0400}&\textbf{1.0708}&\textbf{0.0184}&\textbf{0.0296}& \multirow{2}{*}{9.3400}&8.7101&1.8018&0.0674\\
		4.0mg/mL&Proposed Method& &1.0750&0.0482&0.0337& &\textbf{9.0606}&\textbf{0.4016}&\textbf{0.0299}\\
		\hline
		Iodine+HE Blood&Image-based Method& \multirow{2}{*}{1.0400}&\textbf{1.0594}&\textbf{0.0324}&\textbf{0.0187}& \multirow{2}{*}{8.4700}&6.4151&2.1010&0.2426\\
		2.0mg/mL&Proposed Method& &1.0606&0.0362&0.0198& &\textbf{6.6707}&\textbf{0.8492}&\textbf{0.2124}\\
		\hline
		\multirow{2}{*}{Iodine 15mg/mL}&Image-based Method& \multirow{2}{*}{1.0100}&\textbf{1.2650}&\textbf{0.1241}&\textbf{0.2525}& \multirow{2}{*}{12.4900}&\textbf{12.3921}&0.2761&\textbf{0.0078}\\
		&Proposed Method& &1.2758&0.1302&0.2632& &12.3529&\textbf{0.1122}&0.0110\\
		\hline
		\multirow{2}{*}{Iodine 10mg/mL}&Image-based Method& \multirow{2}{*}{1.0100}&\textbf{1.0991}&\textbf{0.0116}&\textbf{0.0883}& \multirow{2}{*}{11.2500}&11.4684&0.5295&0.0194\\
		&Proposed Method& &1.1096&0.0405&0.0986& &\textbf{11.4268}&\textbf{0.1869}&\textbf{0.0157}\\
		\hline
		\multirow{2}{*}{Iodine 5.0mg/mL}&Image-based Method& \multirow{2}{*}{1.0000}&\textbf{1.0425}&\textbf{0.0107}&\textbf{0.0425}& \multirow{2}{*}{9.6500}&9.0884&1.2136&0.0582\\
		&Proposed Method& &1.0491&0.0382&0.0491& &\textbf{9.2008}&\textbf{0.3066}&\textbf{0.0466}\\
		\hline
		\multirow{2}{*}{Iodine 2.0mg/mL}&Image-based Method& \multirow{2}{*}{1.0000}&\textbf{1.0282}&\textbf{0.0272}&\textbf{0.0282}& \multirow{2}{*}{8.3700}&\textbf{5.8977}&1.8065&\textbf{0.2954}\\
		&Proposed Method& &1.0302&0.0284&0.0302& &5.6042&\textbf{1.1703}&0.3304\\
		\hline
		\multirow{2}{*}{HE Blood 100}&Image-based Method& \multirow{2}{*}{1.1000}&1.1309&0.0779&0.0281& \multirow{2}{*}{7.2600}&\textbf{4.9398}&1.1733&\textbf{0.3196}\\
		&Proposed Method& &\textbf{1.1288}&\textbf{0.0397}&\textbf{0.0262}& &4.0157&\textbf{0.6750}&0.4469\\
		\hline
		\multirow{2}{*}{HE Blood 70}&Image-based Method& \multirow{2}{*}{1.0700}&1.1143&0.0879&0.0414& \multirow{2}{*}{7.3400}&\textbf{5.1408}&1.4103&\textbf{0.2996}\\
		&Proposed Method& &\textbf{1.1122}&\textbf{0.0439}&\textbf{0.0394}& &3.9316&\textbf{0.7626}&0.4644\\
		\hline
		\multirow{2}{*}{HE Blood 40}&Image-based Method& \multirow{2}{*}{1.0300}&1.0795&0.0863&0.0480& \multirow{2}{*}{7.4200}&\textbf{5.1449}&1.3658&\textbf{0.3066}\\
		&Proposed Method& &\textbf{1.0777}&\textbf{0.0432}&\textbf{0.0463}& &3.8966&\textbf{0.8010}&0.4749\\
		\hline
		\multirow{2}{*}{HE Brain}&Image-based Method& \multirow{2}{*}{1.0200}&1.0981&0.1185&0.0765& \multirow{2}{*}{7.4200}&\textbf{5.0935}&1.2363&\textbf{0.3135}\\
		&Proposed Method& &\textbf{1.0957}&\textbf{0.0615}&\textbf{0.0743}& &4.0788&\textbf{0.7598}&0.4503\\
		\hline
		\multirow{2}{*}{True Water}&Image-based Method& \multirow{2}{*}{1.0000}&1.0431&,0.0762&0.0431& \multirow{2}{*}{7.4200}&\textbf{4.9276}&1.1237&\textbf{0.3359}\\
		&Proposed Method& &\textbf{1.0404}&\textbf{0.0380}&\textbf{0.0404}& &4.0875&\textbf{0.5796}&0.4491\\		
		\hline
		\multirow{2}{*}{CT HE Solid Water}&Image-based Method& \multirow{2}{*}{1.0000}&1.0578&0.0954&0.0578& \multirow{2}{*}{7.2400}&\textbf{5.0767}&1.1944&\textbf{0.2988}\\
		&Proposed Method& &\textbf{1.0570}&\textbf{0.0493}&\textbf{0.0570}& &4.0840&\textbf{0.6466}&0.4359\\
		\hline
		\multirow{2}{*}{HE General Adipose}&Image-based Method&\multirow{2}{*}{0.9400}&1.0417&0.1023&0.1082& \multirow{2}{*}{6.4400}&\textbf{4.8484}&0.6123&\textbf{0.2471}\\
		&Proposed Method& &\textbf{1.0411}&\textbf{0.0551}&\textbf{0.1076}& &4.7153&\textbf{0.2434}&0.2678\\
		\bottomrule
	\end{tabular}
	\normalsize
\end{center}
\end{table*}

\begin{table}[ht]
	\caption{Image quality assessments of multi-energy phantom study.}\label{tbl:Gammex_IQA}
	\begin{center}
		\begin{tabular}{r|c|c|c}
			\toprule\vspace{2pt}
			&\multicolumn{3}{c}{Electron Density}\vspace{2pt}\\
			&PSNR&NMAD&SSIM\\
			\midrule
			Image-based Method&36.0653&\textbf{0.0703}&0.9891\\
			Proposed Method&\textbf{36.7651}&0.0706&\textbf{0.9906}\\
			\bottomrule
			&\multicolumn{3}{c}{Effective Atomic Number}\vspace{2pt}\\
			&PSNR&NMAD&SSIM\\
			\midrule
			Image-based Method&29.3859&0.2069&0.9785\\
			Proposed Method&\textbf{29.4807}&\textbf{0.2046}&\textbf{0.9830}\\
			\bottomrule
		\end{tabular}
	\end{center}
\end{table}

\section{DISCUSSIONS}
Conventionally, the estimation of $\rho_e$ and Z maps is based on the decomposed material images from DECT, i.e., the estimation and decomposition are independent processes. However, in this work, we effectively combined the two steps within an optimization model and simultaneously achieved material composition, $\rho_e$ and Z maps. The theoretical superiority lies in two aspects. First, the searched-for material composition, $\rho_e$ and Z maps are efficaciously connected by the proposed model. During the iterative solution process, each map is alternatively updated according to the constrains from the others, i.e., mutual correction. Thus, all the maps are continuously optimized in a consistent manner. Second, we introduced smoothness mechanism for each inverse solution to overcome the illposedness of both decomposition and estimation processes. Thus, the noise can be remarkably suppressed for all the searched-for maps.

In tissue characterization phantom study, results in Tables 1 and 2 demonstrate the improved $\rho_e$ and Z estimation accuracy of the proposed method for most of the materials, especially for high Z materials, such as cortical bone, CaCO3, and inner bone. Although the estimation performance of the proposed method is very similar with the image-based method for low Z materials, i.e.,liver, brain, water, breast, adipose and lung, the standard deviation of the proposed method is much smaller, which reveals the effectiveness of noise suppression and smoothness mechanism of the proposed method. In multi-energy phantom study, although the noise is severer than the previous study, the proposed method can still obtain reasonable $\rho_e$ and Z estimation results (see Tables 4 and 5) compared with the image-based method. Remarkably, the standard deviation of Z estimation by the proposed method is consistently smaller than the image-based method for all the materials. Same conclusion can also be drawn from $\rho_e$ estimation with low Z materials, such as blood, brain, water and adipose. In addition, for both studies, the image quality assessment indexes of PSNR, NMAD, and SSIM shown in Tables 3 and 5 confirm the superior performance of the proposed method in signal to noise ratio and image structural similarity. It is worth noting that both methods cannot accurately estimate all the materials, which is owing to the decomposition rationale. In this inverse problem, we collected two data sets from DECT scan, but we need to estimate $\rho_e$ and Z maps for 13 or 16 materials, which is essentially under-determined. Actually, each map is expressed by the combination of the two selected basis materials. This linear representation mechanism inevitably limits the estimation accuracy of some specific materials. To further overcome the limitation, we will improve and develop the proposed method to a multiple-basis-material based version. In this way, each material can find its optimal expression from the large basis material pool.

Although the proposed method has some superiority and novelty in terms of theoretical model, it still has some limitations confining its direct application. First, there are a lot of parameters need to be determined by an automatic selection strategy. However, up to now, we just experimentally optimized and chose them, but we still did not have some clear rules or indicators. Basically, the regularization and relaxation parameters highly rely on the quality of acquired data and specific structures of searched-for maps. If the collected data is very noisy, then we may prefer larger smoothness parameters. However, if the searched-for maps contain very fine structures, we may sacrifice the smoothness to preserve image details. To improve the application value, we need to provide experimental parameters for common scenarios, propose general rules for parameter optimization, and further develop robust auto-selection strategy.

\section{CONCLUSIONS}
In this work, we propose a one-step iterative estimation method for $\rho_e$ and Z maps with multi-domain gradient $L_0$-norm minimization. The employed gradient $L_0$-norm, as a smoothness regularizer, directly and effectively suppresses noise and reduces artifact in the decomposition domain. The proposed method does not rely on X-ray spectra, i.e., extra spectrum estimations can be avoided. Moreover, it can be accelerated by parallel computing and converge very fast (all the experiments can be finished in 2 minutes with single GPU (NVIDIA GeForce GTX 960M)). Both phantom and patient studies demonstrate the superiority of the proposed method in material-selective reconstruction, noise removal, artifacts reduction and accurate estimation of $\rho_e$ and Z maps.

The clinical meaning of this work lies in more accurate estimation of Z and $\rho_e$, which can be further employed for SPR calculation for proton therapy \cite{harms2020cone}. Moreover, we can synthesize monoenergetic images for dose verification or ROI contrast enhancement. In the future, we will extend this method for multiple material based decomposition, and develop automatic parameter optimization strategies.

\section*{Acknowledgement}
This research is supported in part by the National Cancer Institute of the National Institutes of Health under Award Number R01CA215718.

\bibliographystyle{IEEEtran}
\bibliography{Ref}
\EOD
\end{document}